%% file: FauxThrow.tex
\documentclass[sigconf]{acmart}

\usepackage{booktabs} 

\usepackage[ruled]{algorithm2e} 

\SetAlFnt{\small}
\SetAlCapFnt{\small}
\SetAlCapNameFnt{\small}
\SetAlCapHSkip{0pt}

\usepackage{xcolor}
\usepackage{bbm}
\usepackage{dsfont}

\usepackage{multirow}

\copyrightyear{2022} 
\acmYear{2022} 

\setcopyright{none}
\renewcommand\footnotetextcopyrightpermission[1]{}
\begin{document}
\title{FauxThrow: Exploring the Effects of Incorrect Point of Release in Throwing Motions}

\author{Goksu Yamac}
\affiliation{
  \institution{Trinity College Dublin}
}
\email{yamacg@tcd.ie}

\author{Carol O'Sullivan}
\affiliation{
  \institution{Trinity College Dublin}
  }
\email{Carol.OSullivan@tcd.ie}

\renewcommand\shortauthors{Yamac \& O'Sullivan}

\begin{abstract}
Our aim is to develop a better understanding of how the Point of Release (PoR) of a ball affects the perception of animated throwing motions. We present the results of a perceptual study where participants viewed animations of a virtual human throwing a ball, in which the point of release was modified to be early or late. We found that errors in overarm throws with a late PoR are detected more easily than an early PoR, while the opposite is true for underarm throws. The viewpoint and the distance the ball travels also have an effect on perceived realism. The results of this research can help improve the plausibility of throwing animations in interactive applications such as games or VR.
\end{abstract}

%
%

%
%

\keywords{Perception, motion capture, games, virtual reality}

\maketitle
\pagestyle{plain}

\input{ThrowBody}

\bibliographystyle{ACM-Reference-Format}
\bibliography{FauxThrow}

\end{document}

%% file: ThrowBody.tex
\vspace{-0.25cm}
\section{Introduction}
 
From infancy, humans can recognize when something unexpected happens in the physical world \cite{Baillargeon1998}. When expectations are not met in a virtual setting, such as a game or Virtual Reality (VR) experience, user experience will be impacted. As technologies develop rapidly to support interactive experiences that encompass both the virtual and the real, new problems arise when virtual physical events do not result in expected outcomes. Throwing objects is one of the first physical interactions that an infant learns, and playing ball games is among the most common childhood and sporting activities. Several factors may affect the perception of a thrown virtual ball, e.g., active throwing in VR or controlling a game avatar. The most challenging aspect of simulating virtual throws is accurately detecting the moment when the thrower releases the ball, known as the Point of Release (PoR). 

\begin{figure*}[t]
\begin{center}
\centering
    \includegraphics[width=1.0\linewidth]{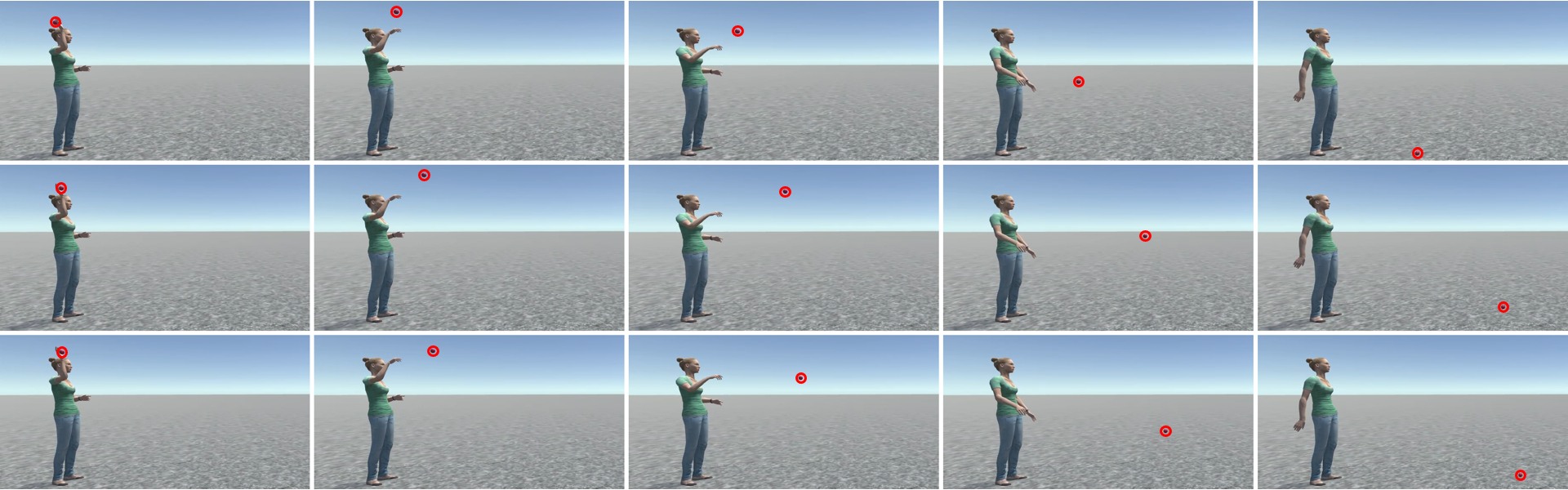}
        \caption{Changes in ball trajectory caused by modifying the Point of Release (PoR) of a throwing motion: early release (t); original trajectory (m); late release (b); for the same five animation frames}
    \label{fig:teaser}
\end{center}
\end{figure*}

Errors in the timing of the PoR can lead to visually disturbing results if the ball's trajectory does not match the viewer's or thrower's expectation (Figure \ref{fig:teaser}). In this paper, we present the first study to examine the effect of release timing on the perception of throwing. Participants watched videos of virtual throwing motions with different levels of PoR timing errors. Their task was to judge whether the motion had been modified or not. The questions we wished to answer from these studies were as follows: Does the \emph{View} from which you observe increase or decrease your ability to detect an anomalous throw, e.g., watching the throw from a \emph{Side View}, as when attending a baseball game or similar; or watching it from a \emph{Front View}, as if throwing the ball oneself?
Is it easier or more difficult to detect an error based on the \emph{Distance} of the throw?
Are different types of throwing motions more or less robust to errors in timing, e.g., \emph{Underarm} vs. \emph{Overarm} throws?
What is the PoR timing \emph{Error} beyond which an anomalous throwing motion can be detected?

\section{Related Work}

O'Sullivan et al. \shortcite{OSullivan2003} explored the factors that affect a user’s perception of a collision in a real-time simulation. In such systems, an approximately accurate result is usually more acceptable than the delay caused by calculating a physically accurate response. They determined that delayed collision responses and angular and momentum distortions of an object's trajectory affect the perceived plausibility of such simulations, and that viewpoint also plays a role. Hoyet et al. \shortcite{Hoyet2012} also found that errors in timing, forces and incorrect angles can reduce the plausibility of physical interactions between virtual people. 

Vicovaro et al. \shortcite{Vicovaro2014} tested user sensitivity to manipulations of overarm and underarm biological throwing animations. Participants perceived shortened underarm throws to be particularly unnatural, and simultaneously modifying the thrower’s motion and the release velocity of the ball did not significantly improve the perceptual plausibility of edited throwing animations. However, editing the angle of release of the ball while leaving the magnitude of release velocity and the motion of the thrower unchanged was found to improve the perceptual plausibility of shortened underarm throws. These studies have motivated the selection of the factors we wish to explore in our experiment.

With respect to active throwing, it was found that people are less accurate in VR than in reality \cite{zindulka2020performance}, mainly due to lower accuracy in distance and height dimensions. This suggests that there were errors in release timing along throw trajectories. Butkus and Ceponis \shortcite{Butkus2019AccuracyOT} found that throwing accuracy in VR increased with distance and that throwing velocity was higher in VR than in reality. 
Covaci et al. \shortcite{Covaci2015} evaluated how effective VR would be for training beginner players to throw a basketball. By tracking a real ball and simulating its continued trajectory within a Virtual Environment, they ensured that the PoR detection and ball trajectory were very accurate. They also explored different viewpoints and found that participants estimated the distance to the basket more accurately from a third-person view. Finally, Faure et al. \shortcite{Faure2020} examined other factors that can affect perception of ball-throwing, such as expertise.

\section{Perception of Point of Release}
Based on the above-mentioned studies, we chose several variables for an experiment exploring viewer sensitivity to errors in the Point of Release (PoR) of a ball during
throwing motions. We hypothesized that the \emph{View} from which participants observe the throwing motions will affect their ability to detect an error, e.g., perhaps certain anomalies, such as angular distortion, would be more visible from the front than from the side. Due to the different velocity properties of long and short throws, our hypothesis was that \emph{Distance} will also play a role, e.g., it could be that errors in slower throws are easier to detect, and that the motion of the \emph{Arm} will also change the dynamics of the throw, as it did in \cite{Vicovaro2014}. We also hypothesized that there will be an asymmetry in early and late PoR timing \emph{Error}, that will vary depending on the other factors.

\begin{figure*}[t]
\begin{center}
     \includegraphics[width=1.0\linewidth]{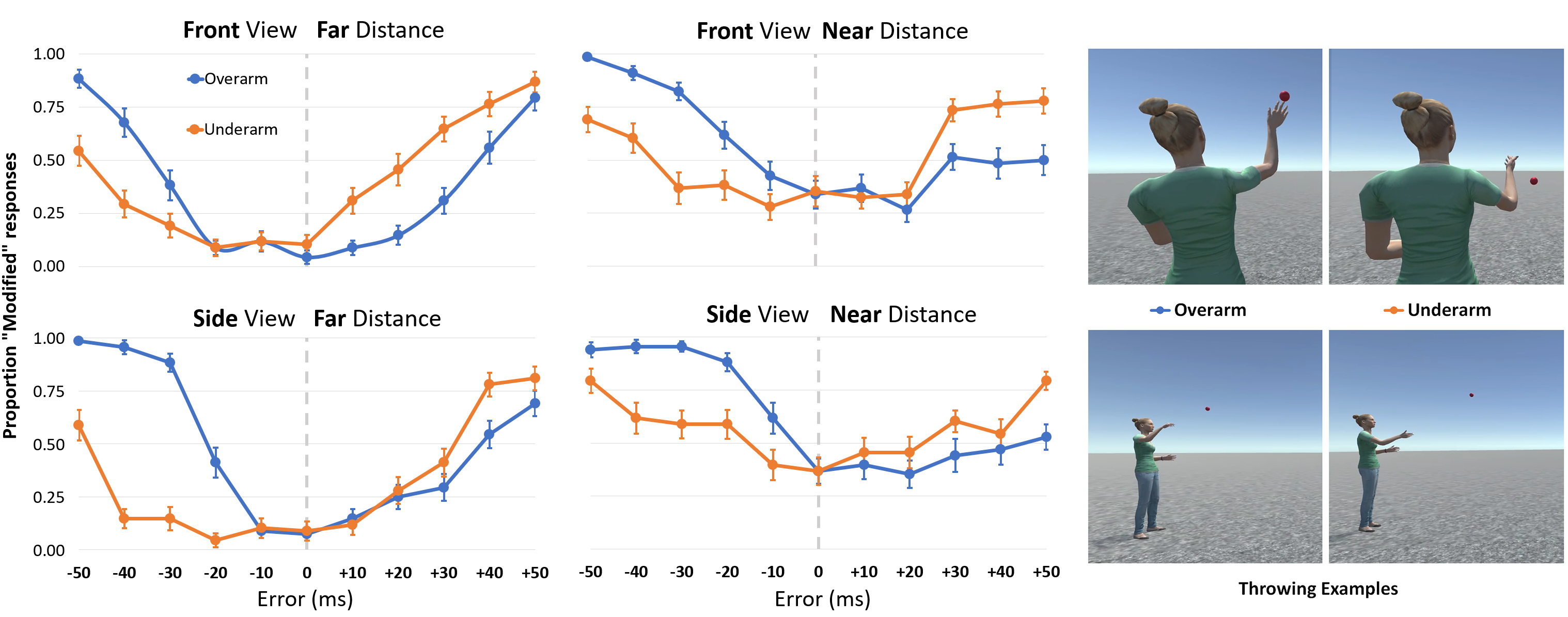}
       \caption{Four-way interaction effect VIEW*DIST*ARM*ERR with means calculated across all participants. Examples frames of throwing motions are shown on the right}
    \label{fig:fourway}
\end{center}
\end{figure*}

\subsection{Stimuli}
We selected the data for our stimuli from a motion capture database that contains the full-body motion of two actors performing a total of 1000 throws to each other, which we planned to use as training data in other work. The data was captured using 21 Vicon\texttrademark{} cameras at 120 frames per second (fps). 
Each actor wore 53 body markers and 20 finger markers (two per finger) and a skeleton was fitted to the marker data by the Vicon software. A 6cm diameter ball was used during the capture, but was not tracked due to interference between finger and ball markers. The actors were instructed to throw as naturally as possible at a range of distances between 2 and 6 metres, without moving from their position. They returned to a neutral posture after each catch before throwing again.

The PoR and Point of Catch (PoC) frames for both actors were calculated in two steps: (i) candidate PoR frames were located, then (ii) a trajectory was fitted between thrower and catcher. In step 1,  frames where the velocity of the hand exceeded a threshold of 1.8 meters per second were found. This threshold was determined by analyzing all the motion data. Flexion at the fingers was checked by examining the mean variance of the second phalange of the index and middle fingers in a window of 20 frames (167ms). Then, as the actors were instructed to remain in position while throwing, hip velocity was checked to ensure that any walking motion was not falsely identified as a throw. When the hip was static and the average finger rotations reached a certain threshold, a candidate PoR was recorded. 

In step 2, we improved on these candidate frames and fitted a ball trajectory between the thrower and the catcher. First, the distance between the actors, and the thrower's hand velocity were used to estimate the airtime for the ball and find a candidate PoC frame. The horizontal velocity of the ball in the air was assumed to be constant. We located search windows of 20 frames around the candidate PoC and candidate PoR frames to calculate the best performing trajectory. In an exhaustive manner, trajectories for the ball were simulated for all potential PoC and PoR frames (ignoring rotational forces and drag). The error is defined as the point on the trajectory with the minimum distance between the ball and the hands of the catcher. The candidate PoR and PoC frames with the trajectory that reached the lowest distance were then selected. 

We selected a total of eight throws (two Overarm-Near, two Underarm-Near, two Overarm-Far, two Underarm-Far) from our dataset to create our stimuli. We sorted all throws based on landing distance and release velocity. Two examples of overarm and underarm throws at two landing distances of similar release velocities were picked based on observed motion neutrality (i.e., no noticeable features such as a lifted left arm or a bent knee) and motion quality (i.e., no retargeting artefacts). Near and Far distances were selected as 1.9 meters and 5 meters, respectively. The PoR and PoC frames were also visualized to ensure that the PoR timing was accurate. 

To create the throwing animations, we retargeted the motions of one female actor to a 3D avatar in Unity\texttrademark{}. We created a ball with a 6cm diameter and attached it to the throwing hand of the avatar. As very high timing precision and control were required, we generated lookup tables to read the position of the ball rather than using Unity's internal event system, which can be unreliable because of changing frame rates. To go beyond the 120 fps precision of the mocap data, quadratic interpolation between positional data points was used to acquire 1000 fps hand position data. The lookup tables contained time-position pairs of the hand. In Unity, the time of an animation clip is represented by normalized time, which takes a value between 0 (start) and 1 (end). We converted this normalized time into the original mocap time of the lookup table. 

At the start of each throw, a new trajectory was generated for the ball based on the delay specified for the current PoR, and a second lookup table was created for the new trajectory. Using these high precision trajectory lookup tables, the ball could be released with any desired delay and the trajectories were not affected by Unity's frame rate. After the ball was positioned using the pre-calculated values in each frame,  Unity's physics system took over to animate the ball when it was close to hitting the ground, in order to simulate a natural bounce (as the catch is not simulated). The clips were then recorded from two different views: the Side view reflects the case where participants observe another person throwing a ball; and the Front view emulates the experience of observing one's own avatar throwing a ball (e.g., as in a VR environment). 

\subsection{Method} 
A total of 34 participants (19M, 15F, ages 18-60+), with a variety of backgrounds and expertise, were recruited via email lists and social media. The experiment was run online using Qualtrics\texttrademark{}. After viewing the instructions and informed consent form, followed by entering some demographical details, participants pressed a key to begin the experiment. No personally identifying information was recorded (i.e., fully anonymous mode was used), and the system ensured that each individual participated only once. The instructions also stated that a display should be used that will allow the motion details to be clearly seen, i.e., a monitor or laptop screen. 

We used a within-subjects design, with independent variables View (Side, Front),  Distance/Dist (Far, Near), Arm (Overarm, Underarm) and 11 different levels of timing Error: $+/-$ 10, 20, 30, 40 and 50 milliseconds, plus the original motion with error 0. This resulted in 2*View $X$ 2*Dist $X$ 2*Arm $X$ 11*Error $=$ 88 combinations, with two repetitions of each using different motion clips, giving 176 video stimuli to be viewed by participants.

Participants viewed either a block of Side view videos first, followed by the Front view block; or vice versa, counterbalanced across all participants. For each block, a set of 8 stimuli was displayed at the start for training purposes, using videos that were not in the set of stimuli. These stimuli consisted of four Original throws (i.e., no PoR delay), from Near and Far distances, and for Overarm and Underarm. Each of the originals was followed by the corresponding Modified throw with an early or late Error. For the modified training examples, we used the most extreme delay of $+/-$ 50 milliseconds. Participants then viewed all 88 videos in that block in random order, and selected a button to indicate whether each throw was Modified or Original.

\section{Results} 
We performed a repeated measures ANalysis Of VAriance with multivariate analysis (ANOVA/MANOVA) to test for main and interaction effects of independent variables VIEW, Distance (DIST), ARM (over or under) and Error (ERR). Newman-Keuls and Bonferroni post-hoc tests were used to check for significant differences. All significant effects are reported in Table \ref{table:stats}.

The main effects show that participants responded Modified more often for Near throws than for Far throws, and for Overarm than for Underarm. The errors were all statistically significantly larger than for the original, except for an early or late release of $+/- 10ms$. However, all the factors interacted with each other, as shown in Figure \ref{fig:fourway}. 
The interaction effects show that, when an Overarm throw had an early PoR error (-50 to -10), it was more often reported as Modified than when the PoR error was late (10 to 50), whereas the opposite was true for Underarm throws. We can also see that participants were much less accurate with their responses for Near throws, where over 30\% of the Original throws (PoR error $= 0$) were mistakenly reported to be Modified. Furthermore, the larger number of Modified responses for Overarm throws than for Underarm is only true for the Side view (VIEW*ARM).


\begin{table}[t]
\begin{center}
\caption{Significant effects ($p<0.05$) for Perception of PoR ANOVA/MANOVA with effect sizes (partial $\eta^{2}$)}
\label{table:stats}

\begin{tabular}{lllll}

	\multicolumn{5}{l}{}\\	
	\small\textbf{Effect}& \textbf{} & \hspace{0.001cm} &\textbf{F-Test}&$\eta^{2}$
	\\\hline\hline
	\multicolumn{5}{l}{\rule{0pt}{3ex}\emph{Main Effects}}\\	\hline
    \multicolumn{2}{l}{{\small DIST}} & \hspace{0.001cm} &$F_{1,33}=36.69, p<0.005$ &0.53\\
    %
    \multicolumn{2}{l}{{\small ARM}}  & \hspace{0.001cm} & $F_{1,33}=18.64, p<0.005$ &0.36\\
    %
    \multicolumn{2}{l}{{\small ERR}} &  \hspace{0.001cm} &$F_{10.330}=97.73, p<0.005$ &0.75\\ \hline
    %
    \multicolumn{3}{l}{\rule{0pt}{3ex}\emph{Two-way Interaction Effects}}\\  	\hline	
    \multicolumn{2}{l}{{\small VIEW $\times$ ARM}} & \hspace{0.001cm} &$F_{1,33}=7.19, p<0.05$ &0.18\\
    %
    \multicolumn{2}{l}{{\small VIEW $\times$ ERR}} & \hspace{0.001cm} &$F_{10,330}=7.44, p<0.005$ &0.18\\
    %
    \multicolumn{2}{l}{{\small DIST $\times$ ERR}} & \hspace{0.001cm} &$F_{10,330}=21.41, p<0.005$ &0.39\\
    %
    \multicolumn{2}{l}{{\small ARM $\times$ ERR}} & \hspace{0.001cm} & $F_{10,330}=42.82, p<0.005$ &0.56\\
    %
    \hline
    \multicolumn{3}{l}{\rule{0pt}{3ex}\emph{Three-way Interaction Effects}}\\  \hline  
    \multicolumn{2}{l}{{\small VIEW $\times$ DIST $\times$ ARM}} & \hspace{0.001cm} & $F_{1,33}=16.16, p<0.005$ &0.33\\
    %
    \multicolumn{2}{l}{{\small VIEW $\times$ DIST $\times$ ERR}} & \hspace{0.001cm} &$F_{10,330}=2.23, p<0.05$ &0.06\\
    %
    \multicolumn{2}{l}{{\small VIEW $\times$ ARM $\times$ ERR}} & \hspace{0.001cm} & $F_{4,80}=2.86, p<0.05$ &0.10\\
    %
    \multicolumn{2}{l}{{\small DIST $\times$ ARM $\times$ ERR}} &\hspace{0.001cm} & $F_{10,330}=3.77, p<0.005$ &0.10\\
    %
    \hline
    \multicolumn{5}{l}{\rule{0pt}{3ex}\emph{Four-way Interaction Effects (See Figure \ref{fig:fourway})}}\\   
    \hline  
    \multicolumn{2}{l}{{\small VIEW $\times$ DIST $\times$ ARM $\times$ ERR}} & \hspace{0.001cm} &$F_{10,330}=3.57, p<0.005$ &0.10\\ 
    %

    \hline\hline \\
\end{tabular}
\end{center}
\end{table}

To explore possible reasons for the perceived accuracy or inaccuracy of these throws, we extracted some error metrics of the ball's motion trajectory for each type of throw and performed a correlation analysis (Table \ref{table:corrPERC}). The Angle metric measures the  horizontal deviation of the angle of the ball's trajectory for each throw with a PoR Error from that of the original throw; the Landing metric gives the deviation in the length of the trajectory from that of the original throw; and the Velocity metric measures the deviation of the  ball's velocity from that of the original throw.

We first tested the correlation of each error metric with the PoR error size. We see, for the Far view only, that significant positive correlations were found for the Angle and Landing errors with the PoR timing error, but not for the Velocity error. Next, we tested for correlations with the participants' error detection ratio, i.e., the proportion of Modified responses. The Front View and Side View entries refer to the average user responses to those same animations. 

We can see that angular deviations are highly correlated with the participant responses, consistent with previous work on collisions \cite{OSullivan2003, Hoyet2012}. This is also evident from simply looking at the stimuli, as a slight twist of the thrower's hand before or after the real PoR can cause very significant angular distortions. The landing position of the ball also had a significant impact on viewers' judgments and, as almost all landing errors involved a shortening of the distance the ball travelled, this is consistent with the results of Hoyet et al. \shortcite{Vicovaro2014}.

\begin{table}[t]

\begin{center}
\caption{Correlations of ball motion errors with participants' error detection ratio. Significant results ($p<0.05$) are highlighted in red}	
\label{table:corrPERC}
											
\begin{tabular}{lllrrr}
& &	&Angle	&Landing	&Velocity\\
	\hline\hline
\multirow{3}{*}{\emph{FAR OVER}} & PoR Error&	&\color{red}0.81	&\color{red}0.98	&0.07\\
&Front View & &\color{red}0.88	&\color{red}0.96	&0.19\\
&Side View &	&\color{red}0.91	&\color{red}0.91	&0.30\\
\hline			

\multirow{3}{*}{\emph{FAR UNDER}} & PoR Error &	&\color{red}0.68	&\color{red}0.76	&0.11\\
&Front View & &\color{red}0.93	&\color{red}0.87	&0.09\\
&Side View &	&\color{red}0.91	&\color{red}0.79	&-0.02\\
\hline

\multirow{3}{*}{\emph{NEAR OVER}} & PoR Error &	&0.25	&0.35	&0.57\\
\hline

\multirow{3}{*}{\emph{NEAR UNDER}} & PoR Error &	&0.54	&0.41	&-0.32\\
&Front View & &\color{red}0.73	&\color{red}0.70	&-0.41\\
&Side View &	&0.43	&0.46	&-0.18\\
\hline\hline
\\
\end{tabular}
\end{center}
\end{table}

\section{Conclusions and Future Work}

In this paper, we presented an experiment on the perception of PoR delays in throwing motion. By manipulating the point of release of motion-captured throwing motions, we have assessed how noticeable different delays are for different distances (Near, Far), different throwing types (Overarm, Underarm) and different views (Front, Side). The results suggest that people are asymmetrically sensitive to early and late delays in overarm and underarm throws. Early release was not as noticeable for underarm throws as it was for overarm throws. Similarly, late releases were less frequently noticed for overarm throws.

However, this was just a first study to explore this problem and we cannot yet generalize from results with this limited number of throws and actors to the wide variety of different throwing types and styles that exist. The presence of a catcher may have affected the motion also. Nevertheless, considering the prevalence of throwing motions in games and VR, we believe that these findings offer valuable insights to guide further studies and the development of interactive applications. We are currently working on developing ML models to accurately predict the PoR for real-time virtual throwing simulations, which incorporate the results of this experiment. Furthermore, we are testing these models in a real-time VR simulation where a small number of simple commodity tracking devices may be used to capture the most important points on the body.

\begin{acks}
This research was supported by Science Foundation Ireland, Grant Agreement 13/RC/2106, at the SFI ADAPT Research Centre, TCD.
\end{acks}